\documentclass[prd,twocolumn]{revtex4}
\usepackage{graphicx, epsfig}
\usepackage{color}
\usepackage{mathrsfs}
\usepackage{bm}

\newcommand{\be}{\begin{equation}}
\newcommand{\ee}{\end{equation}}
\newcommand{\bea}{\begin{eqnarray}}
\newcommand{\eea}{\end{eqnarray}}
\newcommand{\ba}{\begin{eqnarray}}
\newcommand{\ea}{\end{eqnarray}}

\newcommand{\gapp}{\mathrel{\raise.3ex\hbox{$>$}\mkern-14mu
              \lower0.6ex\hbox{$\sim$}}}
\newcommand{\lapp}{\mathrel{\raise.3ex\hbox{$<$}\mkern-14mu
              \lower0.6ex\hbox{$\sim$}}}

\newcommand{\mmbar}{${\rm M\overline{M}}$ }
\newcommand{\sothree}{${\rm SO(3)}$}
\newcommand{\rmo}{r_m}
\newcommand{\rmbar}{r_{\bar m}}

\begin{document}
\title{Monopole-Antimonopole Scattering}
\author{Tanmay Vachaspati}
\affiliation{
Physics Department, Arizona State University, Tempe, AZ 85287, USA.
}

\begin{abstract}
\noindent
We numerically study the head-on scattering of a 't Hooft-Polyakov magnetic monopole and antimonopole
for a wide range of parameters. 
In contrast to the scattering of a $\lambda \phi^4$ kink and antikink in 1+1 dimensions, we find that the 
monopole and antimonopole annihilate even when scattered at relativistic velocities. If the monopole 
and antimonopole have a relative twist, there is a repulsive force between them and they can initially
be reflected. However, in every case we have examined, the reflected monopoles remain bound and 
eventually annihilate. We also calculate the magnetic helicity in the aftermath of monopole-antimonopole 
annihilation and confirm the conversion of relative twist to magnetic helicity as discussed earlier
in the electroweak case.
\end{abstract}

\maketitle

Beautiful results have been obtained on the scattering of monopoles on monopoles 
\cite{Manton:1981mp,AtiyahHitchinbook,MantonSutcliffebook}.
For example, analytical techniques show that head-on collision leads to $90^\circ$ scattering 
for a certain value of the coupling constant (in the so-called Bogomolny-Prasad-Sommerfield (BPS) 
limit) \cite{AtiyahHitchinbook}. 
Monopole-{\it antimonopole} scattering, though, has received less attention, perhaps because 
the process is less amenable to analysis.

A general expectation is that monopole-antimonopole (\mmbar) scattering will lead to their
annihilation and the energy will be dissipated in the form of radiation. However
this is not the result obtained in the analogous process of $Z_2$ kink-antikink scattering in 
1+1 dimensions. Numerical studies of kink-antikink scattering show annihilation at low kinetic energy, 
reflection at higher incoming energy, followed by annihilation at yet higher energies, {\it etc.},
yielding a band structure reminiscent of solutions of the Mathieu equation 
\cite{Campbell:1983xu,Anninos:1991un}. 
One motivation for the present work is to check for chaotic behavior in \mmbar scattering. 

A second motivation for studying \mmbar scattering comes from the recent interest
in the possible existence and detection of a helical inter-galactic magnetic field \cite{Tashiro:2013ita}. 
Early work had speculated on the production of magnetic fields during monopole-antimonopole
annihilation \cite{Vachaspati:1991nm,Vachaspati:1994xc}. The connection to baryogenesis
was made when the electroweak sphaleron solution that mediates baryon number violation 
was interpreted in terms of electroweak \mmbar pairs \cite{Vachaspati:1994ng,Hindmarsh:1993aw}. 
It is crucial for this connection that monopole-antimonopole pairs can have a relative ``twist''
and the  (unstable) electroweak sphaleron solution is really an \mmbar pair that is prevented from annihilating 
by the presence of a twist. In sphaleron decay,  the twist is believed to be the reason that
the resultant magnetic field has non-zero helicity, $h$, defined by
\begin{equation}
h = \int d^3x ~{\bm A}\cdot {\bm B}
\label{helicity}
\end{equation}
where ${\bm A}$ is the electromagnetic gauge potential and ${\bm B}$ is the magnetic field.
In this paper, we will also study the scattering of twisted \mmbar pairs and confirm that
magnetic field helicity originates in the relative twist of the \mmbar.

The results of our investigations are easily summarized: numerical evolution for a 
wide range of \mmbar initial conditions show that untwisted \mmbar scattering always leads to 
annihilation. Thus we do not see any evidence for chaotic behavior similar to that seen in
1+1 dimensions. However, when the monopoles are initially twisted, there is a repulsive force between 
the monopoles. At low velocities, the monopoles slow down or may even reflect back. Yet this
reflection is temporary and soon reversed, and the \mmbar then annihilate. Further, the
annihilation of twisted \mmbar results in the production of a helical magnetic field.

We start out in Sec.~\ref{sec:model} by defining the field theory, describing
the magnetic monopoles and the twisted \mmbar ansatz in which the monopoles are also 
Lorentz boosted. The \mmbar field ansatz will form the initial conditions for the numerical 
evolution described in Sec.~\ref{evolution}, where we also show sample plots of the scattering,
the trajectories of the \mmbar, and the magnetic helicity generated during annihilation. We 
conclude in Sec.~\ref{conclusions}.

\section{SO(3) Model, Monopoles, and \mmbar Ansatz}
\label{sec:model}

\subsection{\sothree model}

The model we study contains an \sothree adjoint scalar and gauge field, $\{\phi^a, W_\mu^a\}$
($a=1,2,3$) with the Lagrangian
\begin{equation}
L = \frac{1}{2} (D_\mu \phi)^a (D^\mu \phi)^a - \frac{1}{4} W^a_{\mu\nu} W^{a \mu\nu}
             - \frac{\lambda}{4} ( \phi^a \phi^a - \eta^2 )^2
\label{lagrangian}
\end{equation}
where,
\begin{equation}
(D_\mu \phi)^a = \partial_\mu\phi^a - i g W^c_\mu (T^c)^{ab} \phi^b 
\end{equation}
and the \sothree generators are $(T^a)^{bc} = -i \epsilon^{abc}$. The gauge field strengths
are defined by
\begin{equation}
W_ {\mu \nu}^a = \partial_\mu W_ \nu^a - \partial_ \nu W_\mu^a + 
  g \epsilon^{abc} W_\mu^b W_ \nu^c.
\label{Wmunu}
\end{equation}

The scalar field equations of motion are
\begin{eqnarray}
\partial_t^2\phi^a &=& 
\partial_i\partial_i \phi^a
+   i g W^{\mu c} (T^c)^{ab}\partial_\mu \phi^b \nonumber \\
&& +   i g W^{\mu c} (T^c)^{ab} (D_\mu \phi)^b 
 - \lambda (\phi^b\phi^b - \eta^2) \phi^a .
 \label{phieq}
\end{eqnarray}
We will work in the Lorenz gauge given by the equation
\begin{equation}
\partial_t W^a_0 = \partial_i W^a_i
\end{equation}
and then the gauge field equations are
\begin{eqnarray}
\partial_t^2 W^a_\mu &=&
\partial_i\partial_i W^a_\mu 
-  g \epsilon^{abc} W^{\nu b}\partial_ \nu W_\mu^c 
- g \epsilon^{abc} W_ \nu^b W^{\nu c} _ {~ \mu} \nonumber \\
&& - g\epsilon^{abc} \phi^b (D_\mu\phi)^c .
\label{Weq}
\end{eqnarray}
By rescaling the coordinates and the fields, as we shall do from now on, we can 
set $g=1$ and $\eta=1$. Then $\lambda$ is the only free parameter left in the 
model. The BPS case is when $\lambda=0$. We will numerically evolve the 15 second
order partial differential equations (PDE) in (\ref{phieq}) and (\ref{Weq}).

The energy density for the model is given by
\begin{eqnarray}
{\cal E} &=&
\frac{1}{2} (D_t \phi)^a (D_t \phi)^a + \frac{1}{2} (D_i \phi)^a (D_i \phi)^a \nonumber \\
&&+ \frac{1}{2} ( W^a_{0i}W^a_{0i} + W^a_{ij}W^a_{ij}) 
+ \frac{\lambda}{4} (\phi^a\phi^a - 1)^2
\end{eqnarray}
where the sum over the repeated index $j$ is restricted to $j >i$.

Once $\phi^a$ acquires its vacuum expectation value, the model contains two massive
gauge fields and one massless gauge field. The massless gauge field is
\begin{equation}
A_\mu = n^a W^a_\mu
\label{Amudefn}
\end{equation}
where $n^a \equiv \phi^a/ \sqrt{\phi^b \phi^b}$ is a unit vector at all spatial points. The field 
strength corresponding to the gauge field $A_\mu$ is defined as \cite{'tHooft:1974qc}
\begin{eqnarray}
A_{\mu\nu} &=& n^a W^a_{\mu\nu} - \epsilon^{abc} n^a (D_\mu n)^b (D_\nu n)^c \nonumber \\
&=& \partial_\mu A_\nu - \partial_\nu A_\mu 
       - \epsilon^{abc} n^a \partial_\mu n^b \partial_\nu n^c .
\label{Amunudefn}
\end{eqnarray}
This field strength definition correspond to the usual Maxwell electric and magnetic field only when 
the magnitude $|\phi|$ is constant. We shall apply them at late times after the monopoles have 
annihilated and when $|\phi|$ is approximately constant and non-zero everywhere.

\subsection{Monopoles}

The monopole solution takes the form
\begin{equation}
\phi^a = P(r) {\hat x}^a
\end{equation}
\begin{equation}
W^a_i = \frac{(1-K(r))}{r} \epsilon^{aij} {\hat x}^j
\end{equation}
where ${\hat x} = {\bm x}/r$ and $r$ is the (rescaled) spherical radial distance centered on 
the monopole. The profile functions $P(r)$, $K(r)$ are not known in closed form except in the 
BPS ($\lambda=0$) case \cite{Prasad:1975kr,Bogomolny:1975de}
\begin{equation}
P_{\rm BPS} (r) = \frac{1}{\tanh(r)}- \frac{1}{r},
\end{equation}
\begin{equation}
K_{\rm BPS} (r) = \frac{r}{\sinh(r)}.
\end{equation}
We will be studying the evolution of monopoles for a range of $\lambda$. A functional form that
reduces to the BPS profile functions for $\lambda=0$ and has the correct asymptotic properties
is
\begin{equation}
P(r) = \frac{1}{\tanh(r)}- (1+m r)\frac{e^{-m r}}{r}
\label{Pprofile}
\end{equation}
\begin{equation}
K(r) = \frac{r}{\sinh(r)}
\label{Kprofile}
\end{equation}
where $m= \sqrt{2\lambda}$ is the scalar particle mass (in $\eta=1$ units).

Next we will need to patch together a monopole and an antimonopole, with a relative twist,
and also boost the monopole and antimonopole towards each other.

\subsection{\mmbar Ansatz}
\label{ansatz}

\begin{figure}
  \includegraphics[height=0.35\textwidth,angle=0]{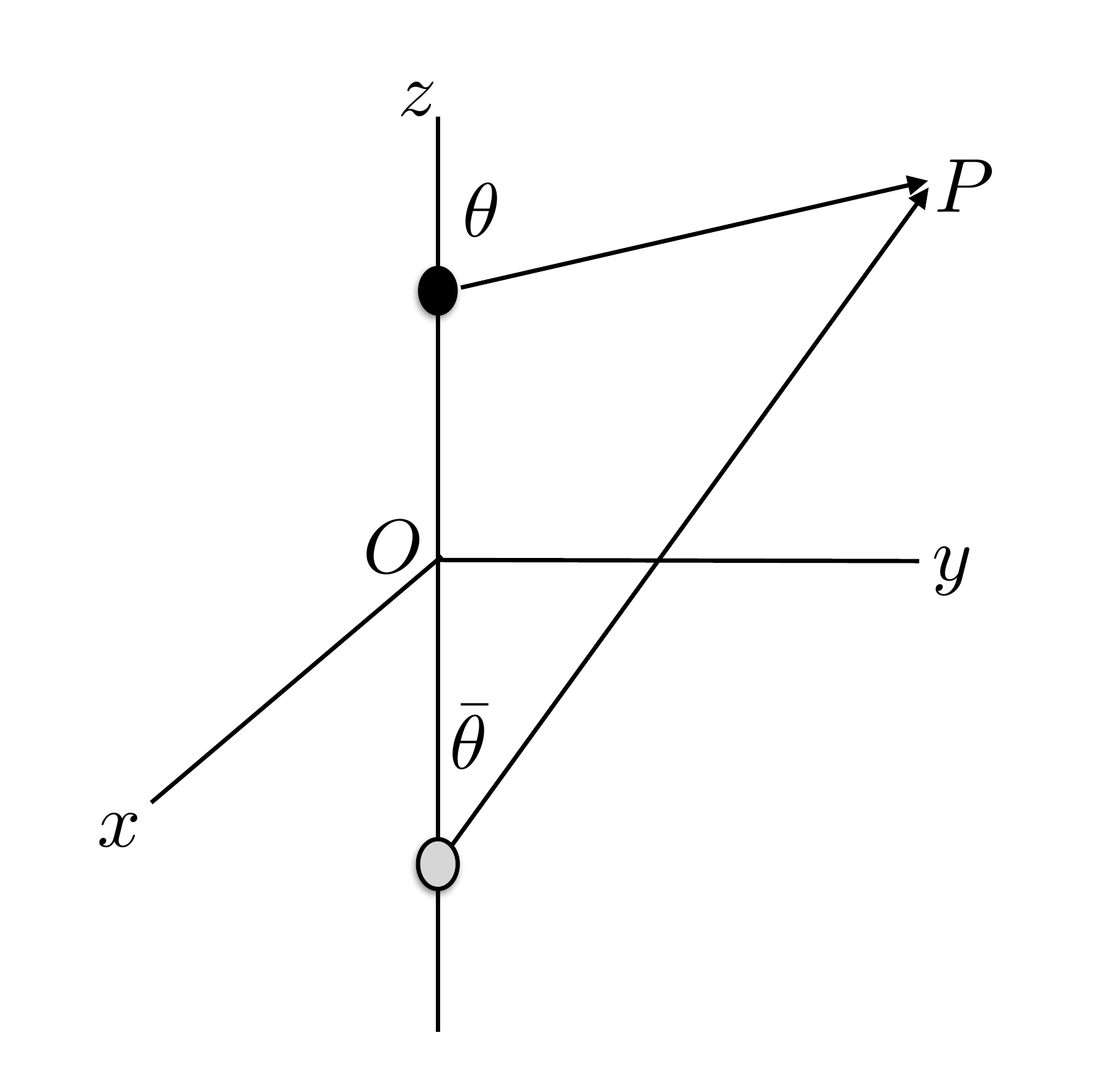}
  \caption{Monopole and antimonopole are chosen to be on the $z-$axis with some initial
  separation $2 z_0$. The spherical angles $\theta$ and ${\bar \theta}$ are defined as shown.
 }
\label{config}
\end{figure}

A twisted monopole-antimonopole ansatz is known in the context of the electroweak model
where the scalar field is an ${\rm SU(2)}$ doublet \cite{Vachaspati:1994ng}. The form is
\begin{equation}
\Phi = \left ( \begin{array}{c}
\sin(\theta /2) \sin({\bar \theta}/2) e^{i\gamma} + \cos(\theta/2)\cos({\bar \theta}/2) \\
\sin(\theta/2) \cos({\bar \theta}/2) e^{i\varphi} - \cos(\theta /2) \sin({\bar \theta}/2) e^{i(\varphi -\gamma)}
\end{array} \right )
\end{equation}
where $\theta$ and ${\bar \theta}$ are the spherical angles centered on the monopole and
antimonopole respectively (see Fig.~\ref{config}), $\varphi$ is the azimuthal angle, and $\gamma$ is 
the twist. A little algebra shows that $\Phi^\dag \Phi =1$.

From $\Phi$, we construct the corresponding unit vector field $n^a$ using
\begin{equation}
n^a = \Phi^\dag \sigma^a \Phi
\end{equation}
where $\sigma^a$ are the Pauli spin matrices. The result, with the replacement 
$\varphi \to \varphi - \gamma/2$ to make the expressions more symmetrical, is
\begin{eqnarray}
n^1 &=& (\sin\theta \cos{\bar \theta} \cos \gamma   - \sin{\bar \theta} \cos\theta)
\cos(\varphi-\gamma/2)  \nonumber \\
&&  - \sin\theta  \sin\gamma \sin(\varphi-\gamma/2) \\
n^2 &=& (\sin\theta \cos{\bar \theta} \cos \gamma   - \sin{\bar \theta} \cos\theta)
\sin(\varphi-\gamma/2)  \nonumber \\
&&  + \sin\theta  \sin\gamma \cos(\varphi-\gamma/2) \\
n^3 &=& \cos\theta \cos{\bar \theta} + \sin\theta \sin{\bar \theta} \cos\gamma
\end{eqnarray}

Close to the monopole, we have ${\bar \theta} \to 0$ and then
\begin{eqnarray}
n^1 &\to& \sin\theta \cos(\varphi+\gamma/2) \\
n^2 &\to& \sin\theta \sin(\varphi+\gamma/2)   \\
n^3 &\to& \cos\theta 
\end{eqnarray}
as we would expect around a monopole. Close to the antimonopole, we have
$\theta \to \pi$ and then
\begin{eqnarray}
n^1 &\to& \sin{\bar \theta} \cos(\varphi-\gamma/2) \\
n^2 &\to& \sin{\bar \theta} \sin(\varphi-\gamma/2)   \\
n^3 &\to& - \cos{\bar \theta} 
\end{eqnarray}
which corresponds to an antimonopole (because of the minus sign in $n^3$). Also
note the relative twist along $\varphi$ of the monopole and antimonopole.

Our ansatz has the nice feature that ${\hat n} \propto {\hat z}$ far away from the \mmbar 
in all directions when the twist vanishes. 
To check this we set $\gamma=0$, ${\bar \theta}\to \theta$ and obtain ${\hat n} = (0,0,1)$.

Now we are ready to write down the scalar field for a twisted monopole-antimonopole pair:
\begin{equation}
\phi^a (x,y,z) = P(\rmo) P(\rmbar) n^a
\end{equation}
where $P(r)$ is the profile function in Eq.~(\ref{Pprofile}) and $\rmo$, $\rmbar$ are the distances of
the spatial point $(x,y,z)$ from the monopole and antimonopole respectively.

At this stage the monopole-antimonopole are at rest. To boost the monopole along the $-z$
direction and the antimonopole along the $+z$ direction we first re-express 
$\rmo$, $\rmbar$ and $n^a$ in Cartesian coordinates
\begin{equation}
\rmo = | {\bm x}-{\bm x}_m | , \ \ 
\rmbar = | {\bm x} -{\bm x}_{\bar m}|
\end{equation}
where ${\bm x}_m =(0,0,z_0)$ and ${\bm x}_{\bar m}=(0,0,-z_0)$ are the locations of the monopole
and the antimonopole respectively. 
The unit vector $n^a$ is also expressed in Cartesian coordinates,
\begin{eqnarray}
\rmo \rmbar n^1 &=& (c x + s y)[ (z+z_0) \cos\gamma - (z-z_0)] \nonumber \\
                                  && - (c y-s x) \rmbar \sin\gamma \\
\rmo \rmbar n^2 &=& (c y - s x)[(z+z_0) \cos\gamma -(z-z_0)] \nonumber \\
                                  && + (c x + s y) \rmbar \sin\gamma \\
\rmo \rmbar n^3 &=& (z-z_0) (z+z_0) + (x^2 + y^2) \cos\gamma 
\end{eqnarray}
where $c \equiv \cos(\gamma/2)$, $s\equiv \sin(\gamma/2)$. Here we have been careful
to distinguish the $(z\pm z_0)$ factors coming from the monopole and antimonopole,
since these will be boosted differently,
\begin{equation}
(z \pm z_0) \to (z \pm z_0)^{(b)} = \gamma_L ((z\pm z_0) \mp v_z t)
\end{equation}
where, $\gamma_L = (1-v_z^2)^{-1/2}$. Note that these boosts also have to be included
in $\rmo$ and $\rmbar$. (We will denote boosted quantities by a $(b)$ superscript.)
Then the scalar fields at $t=0$ for a boosted, twisted monopole-antimonopole pair are:
\begin{equation}
\phi^a (x,y,z) = \left [ P(\rmo^{(b)}) P(\rmbar^{(b)}) n^{{(b)}a} \right ]_{t=0}
\end{equation}
We also need the first time derivative (denoted by an overdot) of the scalar field at $t=0$ and 
this is given by
\begin{equation}
{\dot \phi}^a (x,y,z) = \left [ \partial_t \left ( P(\rmo^{(b)}) P(\rmbar^{(b)}) n^{{(b)}a} \right ) \right ]_{t=0}
\end{equation}
The partial time derivative can be expressed in terms of spatial derivatives as discussed
below.

Now that we have the initial scalar fields, we move on to specify the initial gauge fields. This
is most simply done numerically using the following scheme. We fix the internal space orientation
of the gauge fields by minimizing the covariant derivative. The vacuum solution of $D_\mu {\hat n} =0$ is
\begin{equation}
W^a_\mu  |_{\rm vacuum} = - \epsilon^{abc} {\hat n}^b \partial_\mu {\hat n}^c
\end{equation}
To this we attach profile functions so that the gauge fields are well defined
at the locations of the monopole and antimonopole. So
\begin{eqnarray}
\hskip -1.0 cm 
W^a_\mu  |_{t=0} &=& - \biggl [ 
 (1-K(\rmo^{(b)}))(1-K(\rmbar^{(b)})) \nonumber \\
 && \hskip 2.5 cm \times
       \epsilon^{abc} {\hat n}^{(b)b} \partial_\mu {\hat n}^{(b)c} \biggr ]_{t=0}
\end{eqnarray}
Finally we need the initial time derivative of $W^a_\mu$. We shall treat the spatial and temporal components
differently to enforce the Lorenz gauge condition. For the spatial components, as in the case of the 
scalar field, the time derivative is given by
\begin{eqnarray}
\hskip -0.5 cm 
{\dot W}^a_i  |_{t=0} &=& - \biggl [ \partial_t \biggl ( 
 (1-K(\rmo^{(b)}))(1-K(\rmbar^{(b)})) \nonumber \\
 && \hskip 2.5 cm \times
       \epsilon^{abc} {\hat n}^{(b)b} \partial_i {\hat n}^{(b)c}  \biggr ) \biggr ]_{t=0}
\end{eqnarray}
For the time component of the gauge field, we use the Lorenz gauge condition
\begin{equation}
{\dot W}^a_0  |_{t=0} = [ \partial_i W^a_i ]_{t=0}
\end{equation}

Although the form of the initial conditions is quite involved, they are not too difficult to
implement since temporal derivatives can be related to spatial derivatives using
\begin{equation}
[ (z\pm z_0)^{(b)} ]_{t=0} = \gamma_L (z\pm z_0)
\end{equation}
\begin{equation}
[ \partial_t (z\pm z_0)^{(b)} ]_{t=0} = \mp \gamma_L v_z
\end{equation}
and spatial derivatives can be evaluated numerically.

\section{Evolution}
\label{evolution}

We discretize the $15\times 2$ first-order equations of motion and evolve the system using
the iterated Crank-Nicholson method with two iterations \cite{Teukolsky:1999rm}. Our code
has the novelty that all field theory specific routines are generated symbolically and are then
inserted into a PDE integrating routine. 
We have also implemented absorbing boundary conditions by assuming that all fields
only depend on  $t- r$ where $r$ is the distance from the center of the lattice. For the specific
problem at hand, all the non-trivial dynamics is well within the simulation volume and
the choice of boundary conditions is not crucial.

The initial energy of our ansatz for $\gamma=0$ matches the analytic result for untwisted BPS 
monopoles. During the numerical evolution we have checked energy conservation at the few percent
level at early times, before energy can start leaving the simulation volume. The Lorenz gauge 
condition is also approximately satisfied at all times in the parameter space we have investigated.

\begin{figure}[h]
  \includegraphics[height=0.23\textwidth,angle=0]{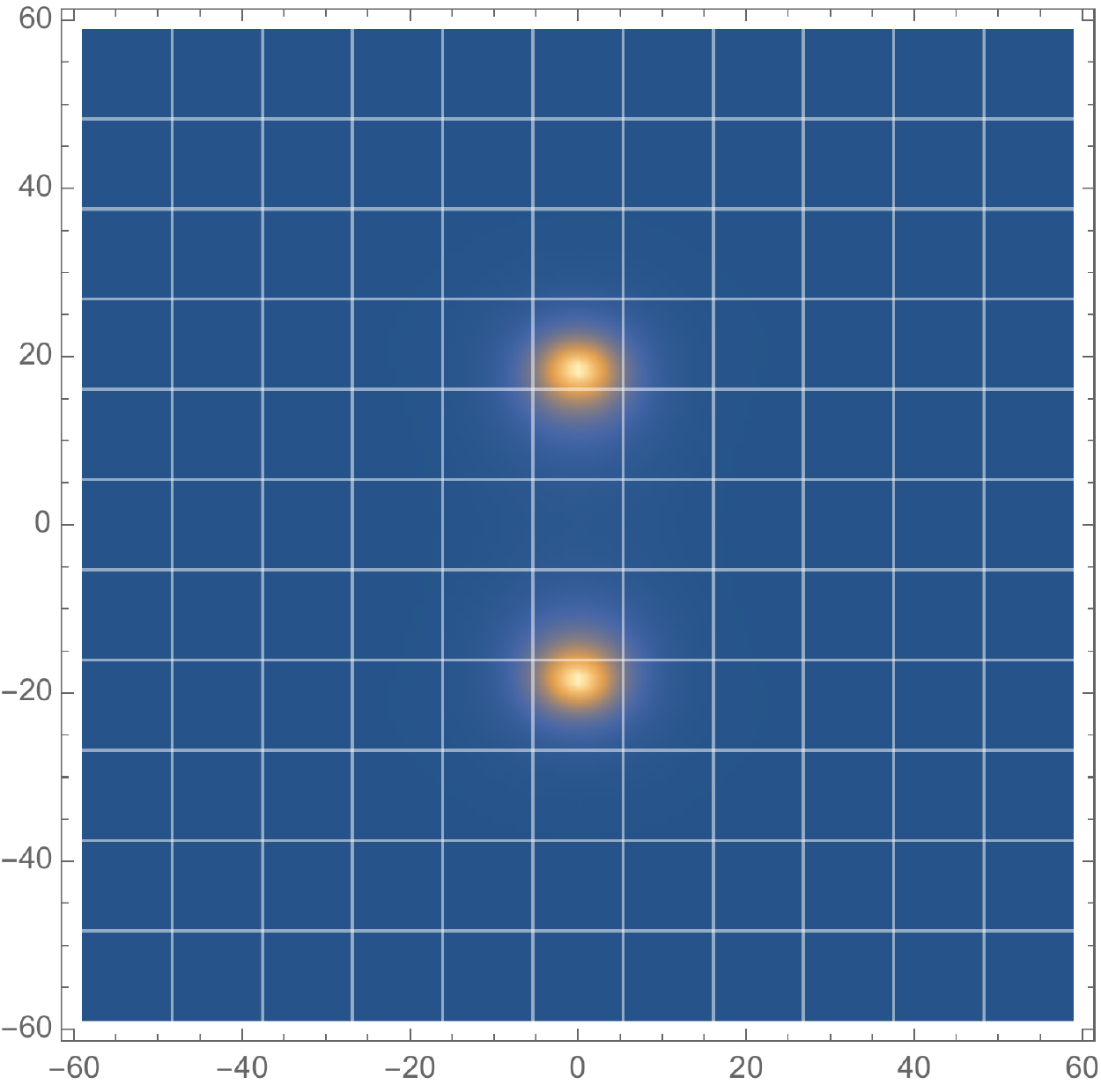}
    \includegraphics[height=0.23\textwidth,angle=0]{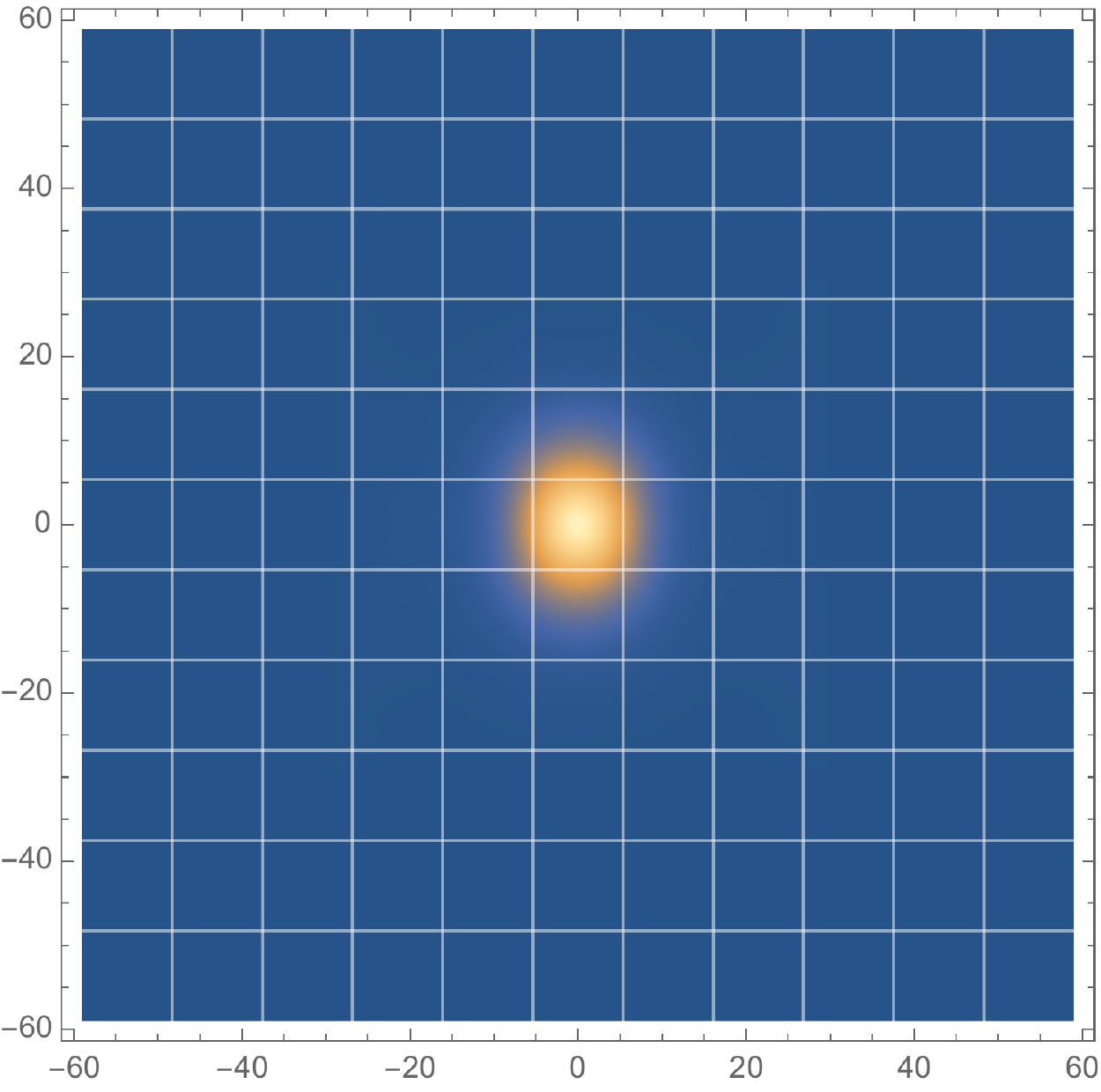}
  \includegraphics[height=0.23\textwidth,angle=0]{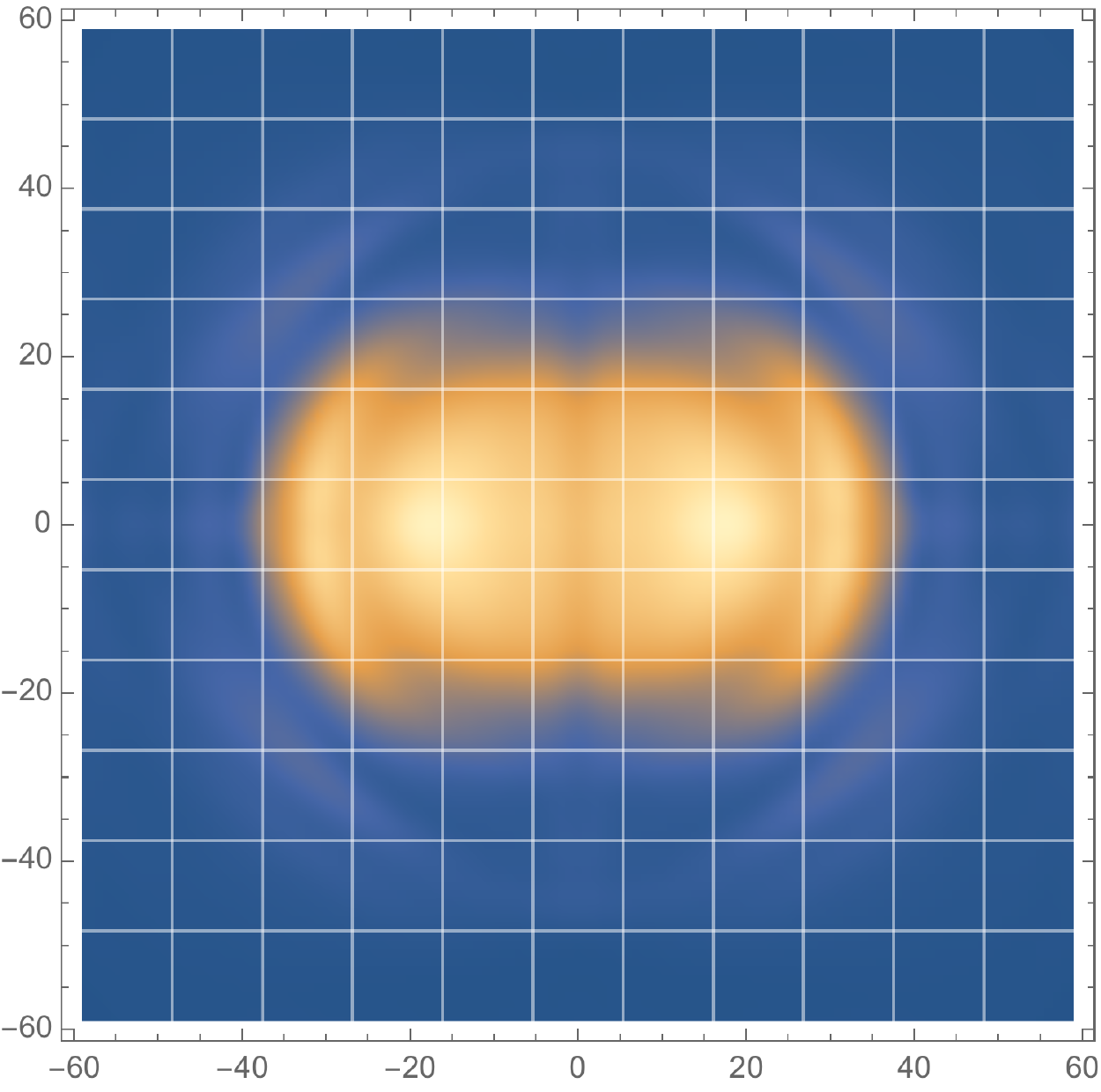}
  \caption{Snapshots of a planar slice of annihilating monopole and antimonopole for $\lambda=1$,
  $\gamma=0$, and $v_z=0.5$. The colors represent energy density.}
\label{snapshots1}
\end{figure}
\begin{figure}
  \includegraphics[height=0.23\textwidth,angle=0]{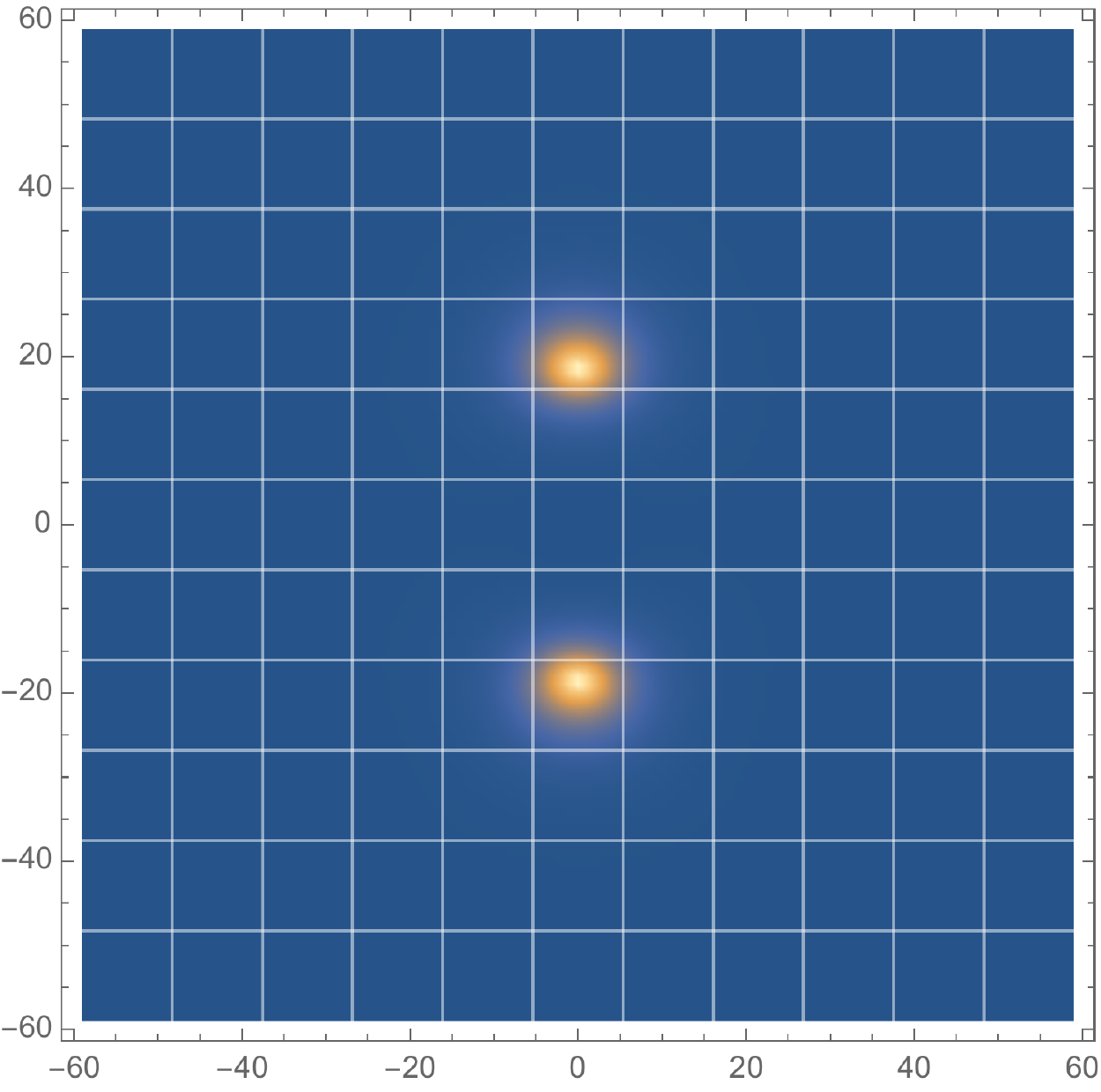}
    \includegraphics[height=0.23\textwidth,angle=0]{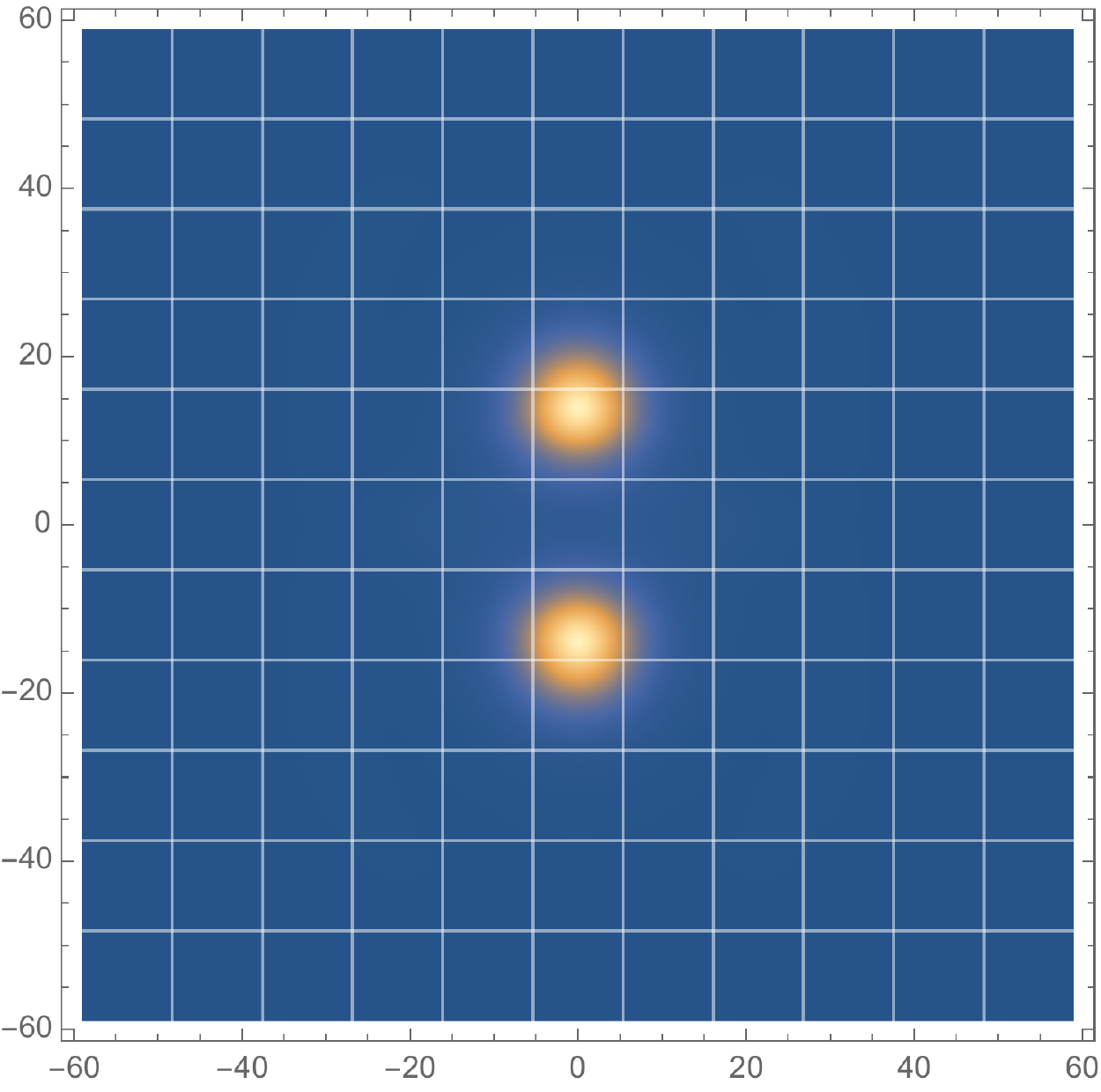}
  \includegraphics[height=0.23\textwidth,angle=0]{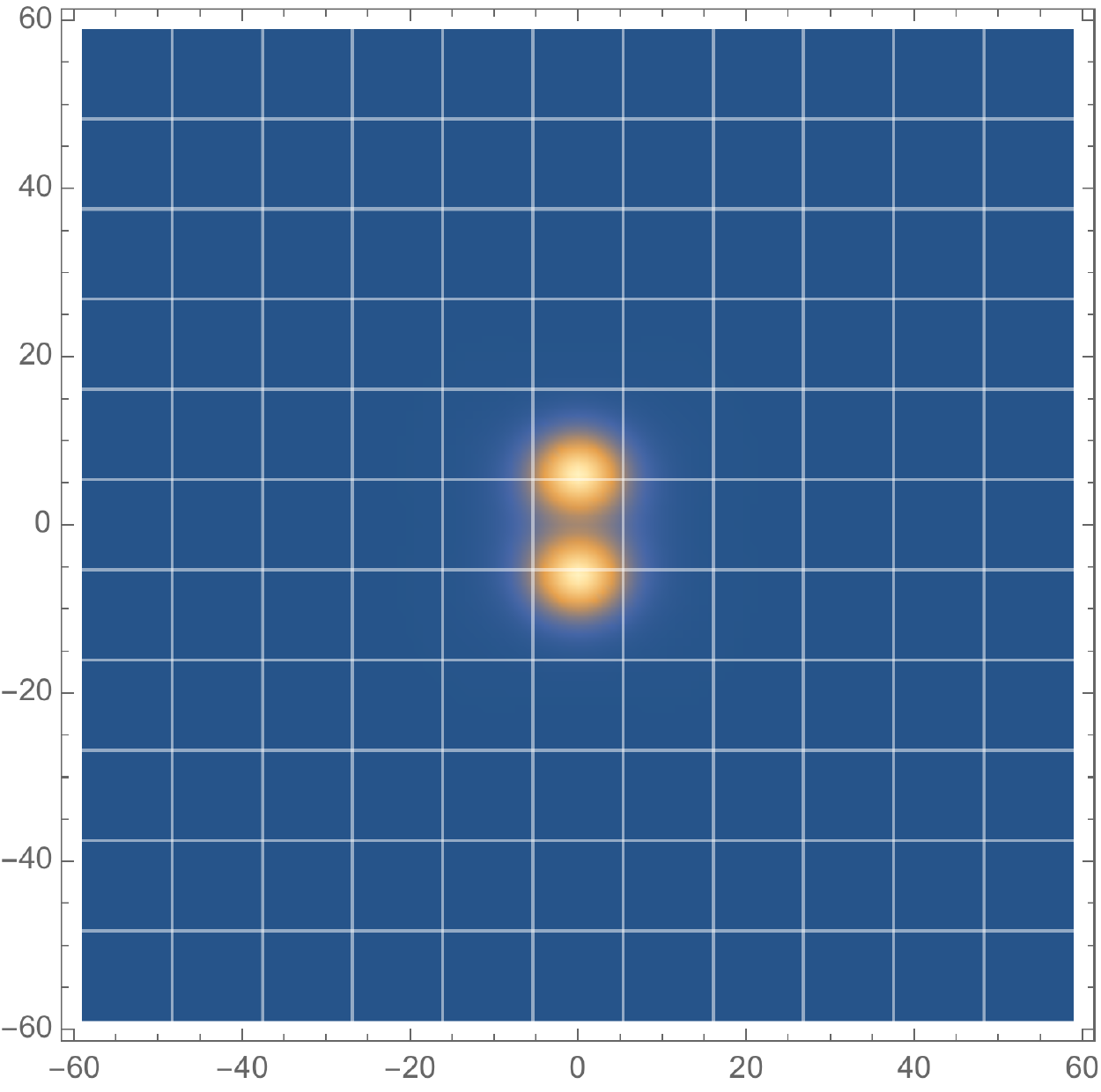}
  \caption{Snapshots of a planar slice of non-annihilating monopole and antimonopole for $\lambda=1$,
  $\gamma=\pi$, and $v_z=0.5$. Except for the twist, all parameters, including snapshot times, are 
  identical to those in Fig.~\ref{snapshots1}. The colors represent energy density. At yet later
  times, the monopoles back-scatter but are still bound and return to annihilate as discussed in the text.}
\label{snapshots2}
\end{figure}

The free parameters in the model are the coupling constant $\lambda$, the boost
velocity $v_z$, and the twist $\gamma$. The initial separation is taken to be 
0.3 times the semi-lattice size plus an offset that ensures that the magnitude of $\phi$
does not vanish on a lattice point at the initial time. (This simplifies some of the
numerics.) We have also chosen $\lambda =1$ 
for our runs, and experimentation with a few other values (including $\lambda=0$)
showed similar results. The initial boost velocity $v_z$ was varied in the interval
$(0.1,0.9)$, and the twist angle was chosen to range from 0 to $2\pi$ in steps of
$\pi/4$. The only runs where we do not explicitly see annihilation until the end of the 
simulation is in the case when $\gamma = \pi$ and for some low values of $v_z$. 
However, even in the cases when the \mmbar do not annihilate, they form a bound 
system and do not escape to infinity. In some cases, we have let the system evolve 
much longer and always found that the \mmbar eventually annihilate.
In Fig.~\ref{snapshots1} we show snapshots of untwisted \mmbar and they simply
come together and annihilate. In Fig.~\ref{snapshots2} we show snapshots of 
twisted ($\gamma = \pi$) \mmbar at the same times as for the untwisted case and
we see that they have not yet annihilated. 

\begin{figure}
  \includegraphics[height=0.25\textwidth,angle=0]{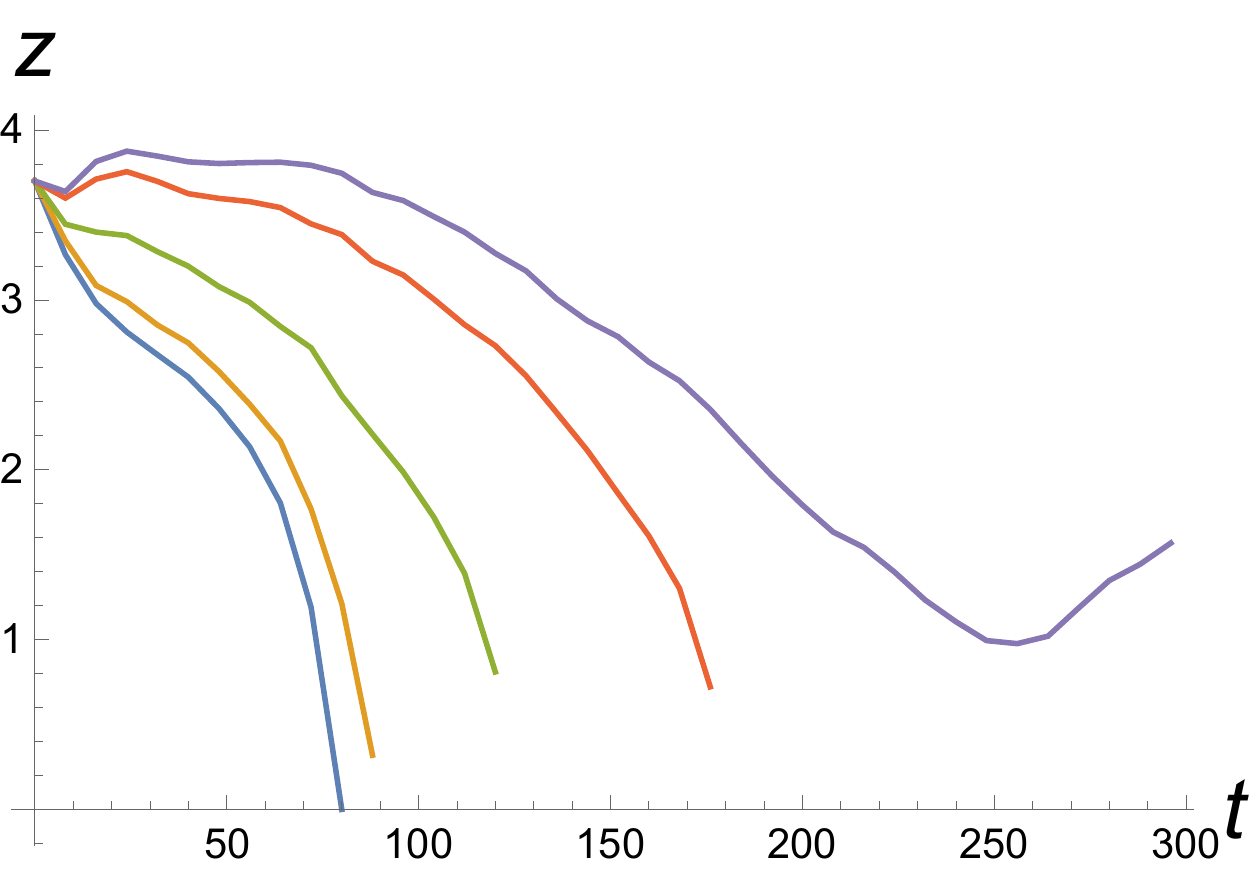}
  \caption{The z-coordinate of the monopole as a function of time for $\lambda=1$,
  $v_z = 0.50$  and $\gamma/(\pi/4) =0, 1, 2, 3, 4$ (curves from left to right).
  The curves terminate once ${\rm min}[\sqrt{\phi^a\phi^a}] \ge 0.25$ (a condition that
  is met after the \mmbar have annihilated) except in the $\gamma=\pi$
  case, when the \mmbar have not annihilated even by the end of the simulation 
  run (300 time steps with $dt=dx/2=0.1$).}
\label{zvstforgamma}
\end{figure}

\begin{figure}
  \includegraphics[height=0.25\textwidth,angle=0]{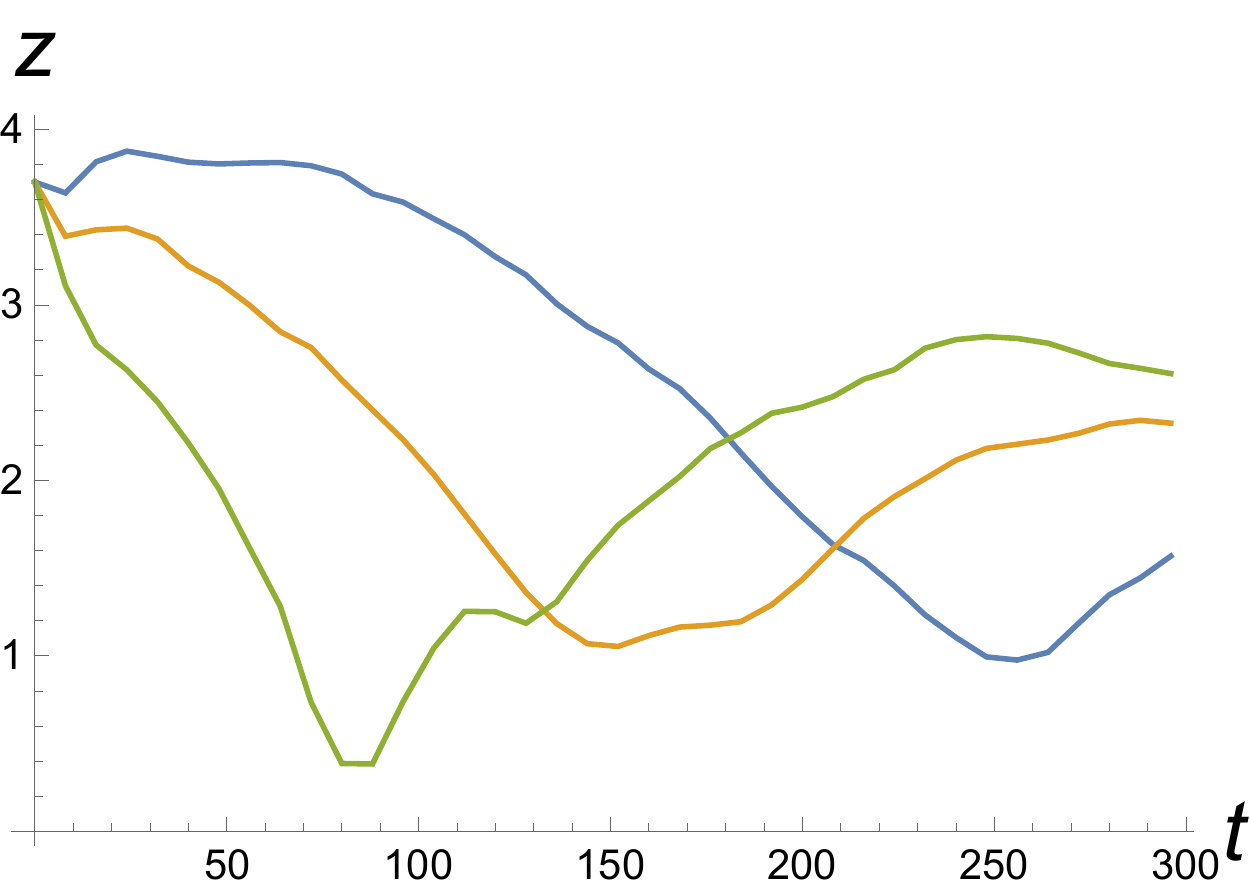}
  \caption{The z-coordinate of the monopole as a function of time for $\lambda=1$,
  $\gamma =\pi$, and $v_z = 0.25, 0.50, 0.75$ (blue, orange and green curves).
  The \mmbar have not annihilated until the end of the simulation.}
\label{zvstforvz}
\end{figure}

We plot the location of the monopole as a function of time for a few sample parameters in 
Figs.~\ref{zvstforgamma} and \ref{zvstforvz}.
The monopole location is defined by the location of the minimum of $\phi^a\phi^a$
over the simulation volume for $z > 0$ provided ${\rm min}[\sqrt{\phi^a\phi^a} ] < 0.25$.
In Fig.~\ref{zvstforgamma}, we hold the velocity fixed at 0.5 and vary the twist from 0 to $\pi$.
(The dynamics for twist of $\gamma$ is the same as that for a twist of $2\pi-\gamma$.) It is clear 
from the plot that the twist slows down the monopole and
can even cause it to bounce back. In Fig.~\ref{zvstforvz} we show $z(t)$ for the monopole
when the twist is held fixed at $\pi$ and $v_z=0.25,0.50,0.75$. Here the bounce back is
very apparent. However, the monopoles are still bound after they bounce back and will
eventually annihilate.
 
\begin{figure}
  \includegraphics[height=0.25\textwidth,angle=0]{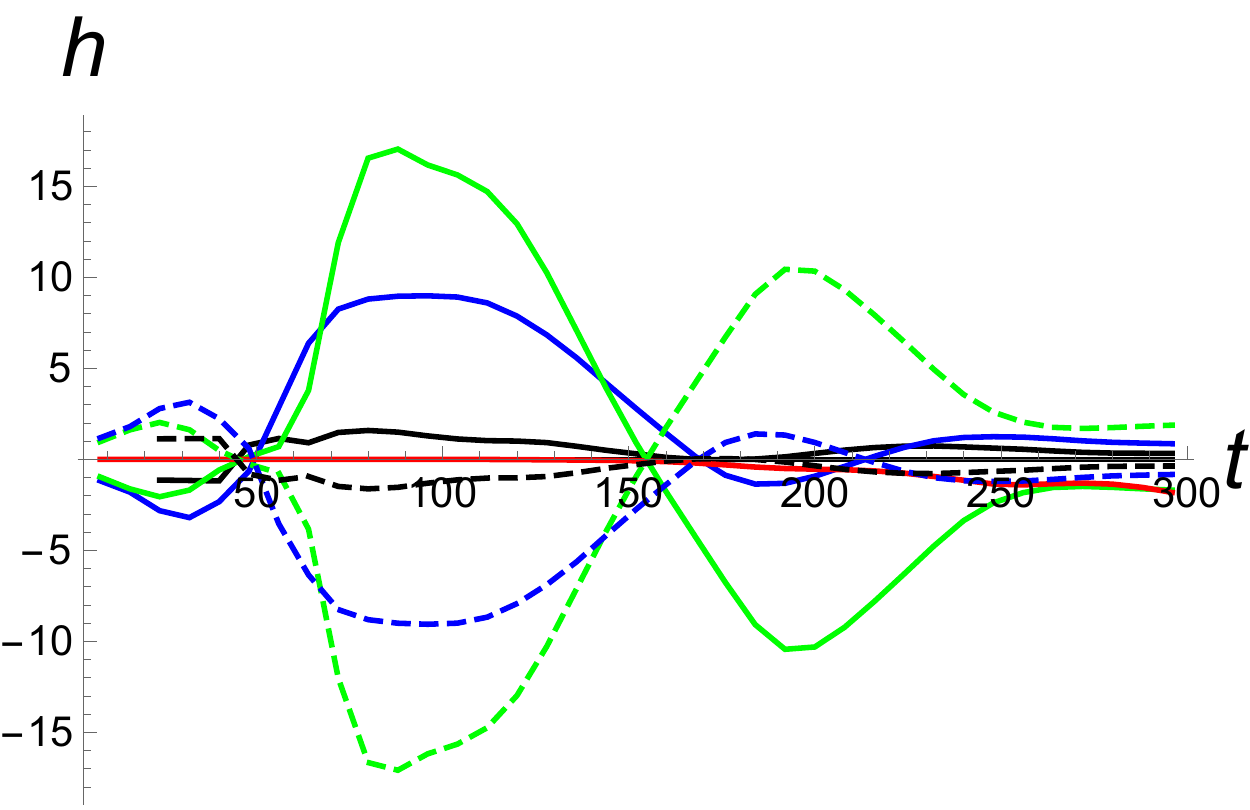}
  \caption{Magnetic helicity in the aftermath of \mmbar annihilation as a function of time
  for $\lambda=1$, $v_z = 0.75$  and $\gamma/(\pi/4) =0,1,2,3,4,5,6,7$. The curves
  for  $\gamma/(\pi/4) =0, 4$ essentially coincide with $h=0$ and are not
  visible. The dashed curves are for $\gamma/(\pi/4) = 5,6,7$ (green, blue, black), and
  mirror the solid curves for $\gamma/(\pi/4) = 3,2,1$ (green, blue, black). This
  shows that $h(\gamma) = - h(2\pi-\gamma)$.}
\label{hvstforgamma}
\end{figure}

The untwisting and annihilation of the \mmbar is expected to radiate magnetic fields
that are helical \cite{Copi:2008he,Chu:2011tx}. To test this expectation, we have calculated 
the helicity defined in Eq.~(\ref{helicity}) using (\ref{Amudefn}) and (\ref{Amunudefn}).
The plot of the magnetic helicity as a function of time is shown in Fig.~\ref{hvstforgamma}
where we hold the velocity fixed at 0.75 and vary the twist. The plot shows that
the helicity vanishes if there is no twist ($\gamma=0$). Also, we see that
the $h(\gamma)=-h(2\pi-\gamma)$, and the helicity vanishes in the case $\gamma=\pi$ 
(though in this case the \mmbar survive until the end of the simulation). These
observations can be understood if the helicity is due to the untwisting motion of
the \mmbar. If $\gamma < \pi$, the \mmbar untwist in one direction and then annihilate,
while if $\gamma > \pi$, the \mmbar untwist in the other direction so that $\gamma \to
2\pi$. The opposite senses of untwisting lead to the production of magnetic fields
with opposite helicity. The value $\gamma=\pi$ is an unstable point where the 
\mmbar are unable to decide which way to untwist. Eventually numerical instabilities will 
cause untwisting in one way or the other. In Fig.~\ref{hvstforgamma} we also observe oscillations 
in the magnetic helicity, suggesting that there may be oscillations in the twist.

\begin{figure}
  \includegraphics[height=0.25\textwidth,angle=0]{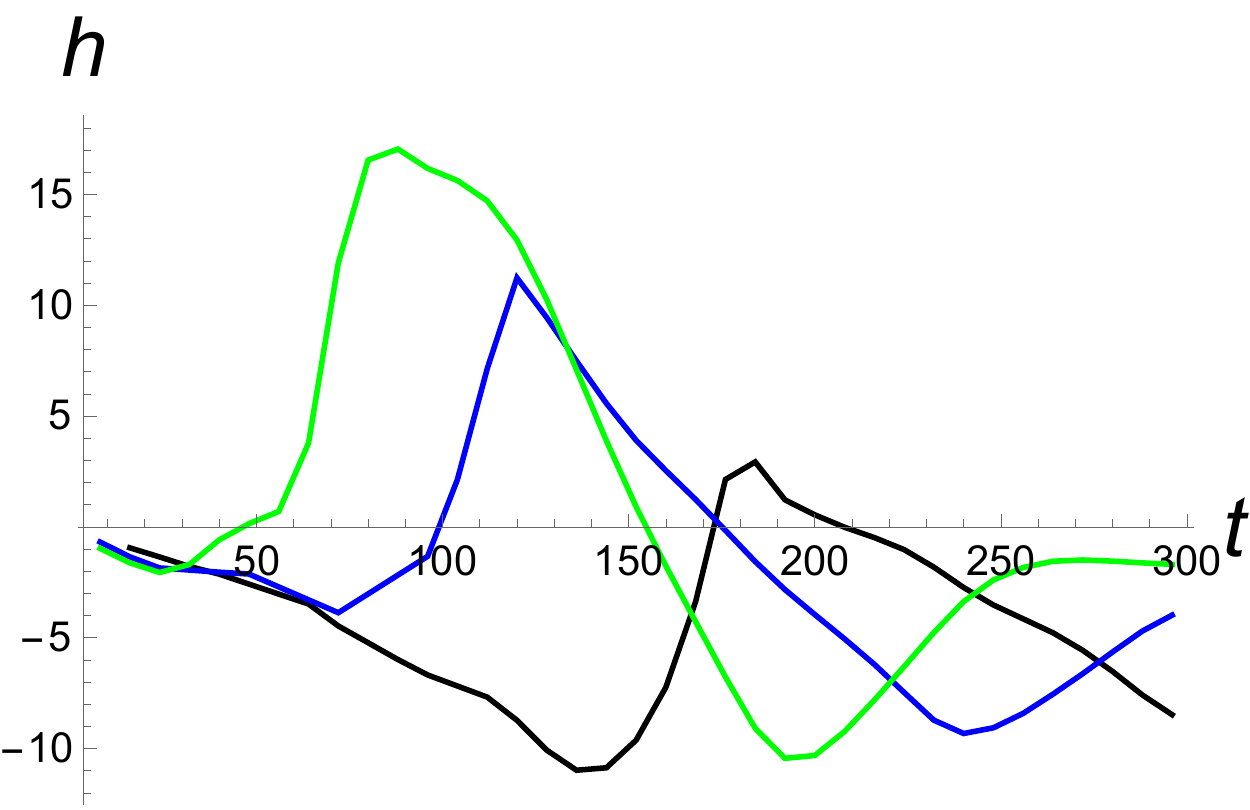}
  \caption{Magnetic helicity in the aftermath of \mmbar annihilation as a function of time
  for $\lambda=1$,  $\gamma = 3\pi /4$, and $v_z = 0.25, 0.50, 0.75$ (black, blue and green
  curves).}
\label{hvstforvz}
\end{figure}

In Fig.~\ref{hvstforvz} we plot the magnetic helicity for $\lambda=1$, $\gamma =3\pi/4$
and for $v_z=0.25,0.50,0.75$. The plots are similar in shape but shifted to earlier
times for higher velocities. This can be understood because the \mmbar scatter
at earlier times for higher velocities.

\section{Conclusions}
\label{conclusions}

We have studied \mmbar scattering by numerical methods. Part of the challenge was to 
devise initial conditions that are suitable to describe boosted and twisted \mmbar. Our 
ansatz for initial conditions are given in Sec.~\ref{ansatz} but there may be other choices. 

The numerical evolution of \mmbar shows that, unlike the scattering of kinks in 1+1 dimensions, 
\mmbar scattering is not chaotic, as the \mmbar are always found to annihilate over the wide
range of parameters we have investigated. A twist in the initial conditions produces a repulsive 
force between the monopole and antimonopole that can have an important effect on the scattering 
dynamics. An interpretation of our results is that, as the \mmbar approach each other, they also
tend to untwist. The untwisting dynamics is damped due to radiation and eventually the \mmbar
can annihilate. However, damping of the untwisting dynamics leads to the production of helical 
magnetic fields and the sign of the magnetic helicity is related to the direction of untwisting.

\acknowledgements

I am grateful to Nick Manton and Daniele Steer for helpful comments.
The computations were done on the A2C2 Saguaro Cluster at ASU. 
This work was supported by the U.S. Department
of Energy, Office of High Energy Physics, under Award No. DE-SC0013605
at ASU.

\end{document}